\documentclass[12pt]{iopart}

\expandafter\let\csname equation*\endcsname\relax
\expandafter\let\csname endequation*\endcsname\relax
\usepackage{amsmath, iopams}
\usepackage{slashed}

\usepackage[margin=0.5cm]{caption}
\usepackage{tikz}\usetikzlibrary{positioning}
\usepackage{graphics}
\usepackage{subfigure}
\newtheorem{definition}{Definition}

\newtheorem{theorem}{Theorem}

\usepackage{cancel}
\usepackage{amsfonts,amssymb}
\def\rvarphi{\raise 2pt\hbox{$\varphi$}}
\usepackage{graphicx}
\usepackage{float}
\usepackage{subfig}
\usepackage{overpic}

\usepackage{import}

\begin{document}

\title{Geometric Inequality for Axisymmetric Black Holes With Angular Momentum}

\author{Xuefeng Feng$^{1,2,3,4}$\footnote{\label{note1} Joint first author.}, Ruodi Yan$^{5}$\footnote{\label{note1} Joint first author.}, Sijie Gao$^{5}$, Yun-Kau Lau$^{6}$\footnote{Corresponding author, email: lau@amss.ac.cn.}, Shing-Tung Yau$^{3,4}$\footnote{Corresponding author, email: styau@tsinghua.edu.cn.}}

\address{$^1$Beijing Institute of Mathematical Sciences and Applications, Yanqihu, Huairou District, Beijing, 101408, China}
\address{$^2$Yau Mathematical Sciences Center, Tsinghua University, Beijing, 100084, China}
\address{$^3$Center for Mathematics and Interdisciplinary Sciences, Fudan University, Shanghai, 200433, China}
\address{$^4$Shanghai Institute for Mathematics and Interdisciplinary Sciences (SIMIS), Shanghai, 200433, China}
\address{$^5$ Department of physics, Beijing Normal University, Beijing 100875, China}
\address{$^6$  Institute of Applied Mathematics, Morningside Center of Mathematics, LSSC, Academy of Mathematics and System Science, Chinese Academy of Sciences, 55, Zhongguancun Donglu, Beijing, 100190, China.}

\begin{abstract}
In an effort to understand the Penrose inequality for black holes with angular momentum, an axisymmetric, vacuum, asymptotically Euclidean initial data set subject to certain quasi-stationary conditions is  considered for a case study. A new geometric definition of angular velocity of a rotating black hole is defined in terms of the momentum constraint, without any reference to a stationary Killing vector field. The momentum constraint is then shown to be equivalent to the dynamics of a two-dimensional steady compressible fluid flow governed by a quasi-conformal mapping.  In terms of spinors, a generalised first law for rotating black holes (possibly with multi-connected horizon located along the symmetry axis) is then proven and may be regarded as a Penrose-type inequality for black holes with angular momentum.

\end{abstract}

\maketitle

\section{Introduction}

For an initial data set consisting of black holes, the Penrose inequality asserts that the norm of the ADM energy-momentum four vector is bounded below by the areal radius of the outermost trapped surfaces \cite{penrose1982some}. So far the inequality is proven only in the time symmetric case \cite{huisken2001inverse,bray2001proof}, the problem remains outstanding when ADM momentum is taken into consideration. The inequality affords a possible further generalisation to rotating black holes with angular momentum of the trapped surfaces included in some way. This was first suggested in the work of Christodoulou \cite{christodoulou1970reversible} and attempt had been made to understand this more general form of Penrose inequality  for axisymmetric initial data set
when angular momentum of a trapped surface is defined in terms of a Komar integral  \cite{dain2006variational,dain2008proof,dain2012geometric}. There exist several reviews on the Penrose inequality \cite{mars2009present,bray2004penrose}.

Motivated by an effort to understand the outstanding Penrose inequality for black holes with angular momentum \cite{dain2012geometric,anglada2018penrose}, the present work aims to study the problem within the specific context of an axisymmetric, quasi-stationary  initial data set \cite{bardeen_variational_1970,dain2006variational}, as a small step towards the understanding of the Penrose inequality for rotating black holes, first put forward in \cite{christodoulou1970reversible}.

In contrast to previous works, we will not plunge straightaway into the proof of a theorem. Instead,  motivated by the Kerr metric, we will first attempt to understand the angular velocity of a spinning black hole in a more general and geometric context, without any reference to a Killing
vector field. We shall give a more general definition of angular velocity on an initial data slice in terms of the constraint equations. It will be shown that the angular velocity  of a spinning black hole may be regarded as a potential that generates the momentum constraint. In doing so, a quasi-conformal continuity equation for two-dimensional steady compressible fluid flow is proven to be equivalent to the momentum constraint and a stream function dual to the angular velocity may be defined in a sense to be given in what follows. A two-dimensional compressible fluid picture then emerges naturally for the dynamics of the momentum constraint.

Given the geometric definition of angular velocity, together with spinorial technique \cite{witten_new_1981} to prove a Penrose-type inequality for the ADM mass, an inequality bounding the ADM mass in terms of the areal radius and the angular momentum of the horizon (possibly multi-connected) may be worked out. 
The inequality may also be regarded  as a generalised first law of black hole mechanics for a black hole yet to settle down to its stationary state and at the same time a Penrose-type inequality for black holes with angular momentum. 

For reference purpose, the precise statement regarding Penrose inequality for black holes with spin was first stated in \cite{christodoulou1970reversible}. Attempt to prove a positive energy theorem for rotating black holes was first made in \cite{shaw1985witten}. This was later followed up in \cite{zhang1999angular}. Lately there have also been attempts in this direction \cite{khuri2019penrose}.

The paper is structured as follows. After some preliminaries and notations in Section \ref{sec:pre}, in Section \ref{sec:ang}, we work out a more general definition of angular velocity of a rotating black hole in terms of the constraint equations and discuss further its geometry. A two-dimensional compressible fluid picture for the momentum constraint then naturally emerges. In Section \ref{sec:theorem}, a generalised first law of black hole mechanics is proven using the technique of spinors. This is to be followed by some concluding remarks in Section \ref{sec:conclude}.

\section{Preliminaries and Notations}\label{sec:pre}

Some background materials  relevant to the present work will be briefly described in this section. Let $(M,g_{ab})$ be a smooth, connected four-dimensional spacetime manifold with Lorentzian metric signature 
$(+,\,-,\,-,\,-)$. Let $N$ be an orientable, complete Riemannian three manifold  identically embedded in $(M, g_{ab})$  so that when restricted to $N$,
\begin{eqnarray}
g_{ab}=\tau_a\,\tau_b-h_{ab}\,,
\end{eqnarray}
where $h_{ab}$ is a smooth Riemannian metric of $N$ and $\tau^a$ is the unit timelike normal of $N$ in $M$. It is standard to denote the second fundamental for the embedding of
$N$ in $M$ by $K_{ab}$. 

An initial data set $(N,h_{ab}, K_{ab})$ in $M$ is assumed to be asymptotically Euclidean in the
standard sense that, in
 the complement of some compact set,
in $N$, 
\begin{eqnarray}
    h_{ab}=e_{ab}+\,\mathcal{O}(1/r),
\end{eqnarray}
where $e_{ab}$ is an Euclidean metric and 
\begin{eqnarray}
     \partial
h_{ab}=\mathcal{O}(1/r^{2}),
 \quad \partial^2 h_{ab}=\mathcal{O}(1/r^{3}),\nonumber\\
 K_{ab}=\mathcal{O}(1/r^{2}),\quad\,\, \partial K_{ab}=\mathcal{O}(1/r^{3})\,,
\end{eqnarray}
where $r$ is the standard radial parameter defined in terms of the
Cartesian coordinates near infinity. 

Denote by $\partial N$ the inner boundary of $N$. $\partial N$ is
assumed to  consist of connected components $S_i$, $i=0,1,\cdots n$
with each $S_i$  a smooth spherical two surface. Let $\gamma_{ab}$
and $p\,$ be respectively the two metric  and
 the mean curvature  of $S_i$ defined with respect to the future (past) pointing null normal.
  Then
\begin{eqnarray}\label{trapped}
  \gamma^{ab}K_{ab}\pm p=0
\end{eqnarray}
characterise $S_i$ as a future $(+)$ and past $(-)$ marginally trapped surface. The stability of $S_i$ is defined in the standard way in terms of the second variation of the area functional along the normal of $S_i$ \cite{newman1987topology}.

\subsection{Quasi-stationary initial data set}
\begin{figure}[h]
\centering
\includegraphics[width=0.2\textwidth]{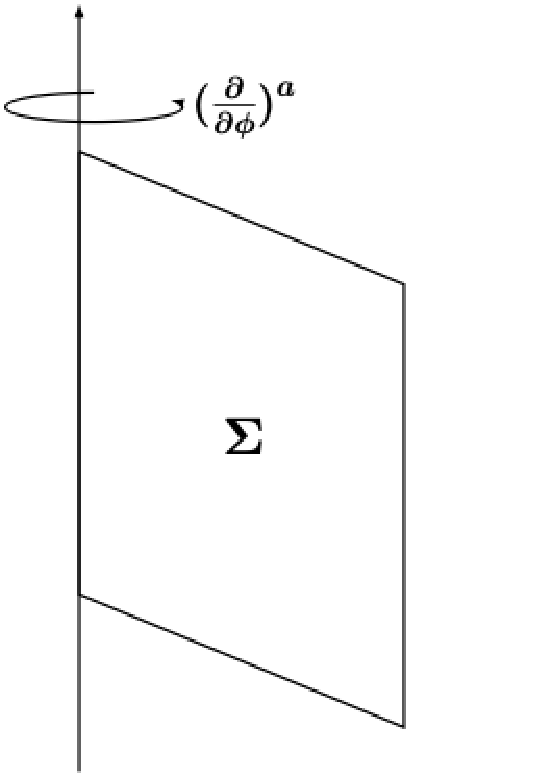}
\caption{ Three manifold $N$  generated by Lie dragging of $\Sigma$ along the integral curves of $\phi^a$.}
\label{fig:rotation}
\end{figure}

Throughout this work we shall further restrict $(N, h_{ab},K_{ab})$ to be axisymmetric and quasi-stationary \cite{bardeen_variational_1970} (or $(t,\phi)$-symmetric) which is a particular case of the Brill data considered in \cite{dain2006variational}. Assume there exists a circle action on $N$ so that $N=\Sigma\times S^1$. $\Sigma$ is diffeomophic to a copy of the half plane
$R^2_+=\{(x,y)\in R^2, x\ge 0\}$, with the symmetry axis of $\phi^a$ as the boundary (see Fig.~\ref{fig:rotation}). The axisymmetric Killing vector $\phi^a$ that generates the circle action is orthogonal to $\Sigma$ 
with 
\begin{eqnarray}
\label{lie}    
    \mathcal{L}_\phi h_{ab}=\mathcal{L}_\phi K_{ab}=0\,,
\end{eqnarray}
where $\mathcal{L}_\phi$ denotes the Lie derivative with respect to $\phi^{a}$.
The three metric of $N$ may be written as 
\begin{eqnarray}
    h_{ab}=P_{ab}+\eta_a\eta_b\,,d
\end{eqnarray}
where $P_{ab}$ is the two metric of $\Sigma$ and $\eta^a=X^{-\frac{1}{2}}\phi^a$ with $X=\phi^a\phi_a$. $K_{ab}$
is further subject to 
\begin{eqnarray}
K_{ab}P^a{}_lP^b{}_m =0\,,\quad K_{ab}\eta^a\eta^b=0\,.\label{extrinsic}
\end{eqnarray}

The axisymmetric structure also means that the future (past) mariginally trapped surfaces $S_i,\,i=1,\cdots n$ 
are generated by simple curves $\gamma_i\subset \Sigma,\, i=1\cdots n$ with end points located at the symmetry axis of 
$\phi^a$. $S_i$ is obtained by Lie dragging of $\gamma_i$ along the circular integral curves of $\phi^a$ (see Fig.~\ref{fig:rotation}). We proceed to let $D$ and $\cancel{\nabla}$ be the connection with respect to the metric $h_{ab}$ and $P_{ab}$.

The constraint equations for $(h_{ab}, K_{ab}, N)$ are also subject to the vacuum Einstein field equations and the maximal slicing condition so that 
\begin{eqnarray}\label{Constraint_Equations_1}
D_a K^{ab}  = 0\,, \\
R - K_{ab} K^{ab} = 0\,,
\end{eqnarray}
where $R$ is the scalar curvature of $(N,h_{ab})$. 

Though the geometry of an axisymmetric initial data set described above resembles  locally to that of the constant time slice in the Kerr metric,  the initial data set is however not as restrictive as it may seem, as globally the three geometry may admit multi-connected horizon, with the initial data of  binary spinning black holes as a particular example. Unlike the majority of previous works, we will adopt a moving frame approach in our subsequent analysis. 

\vskip 20pt
\subsection{Background in spinors}

Spinors will also be employed in the present work. Denote by $\tau^{AA'}$, the timelike unit normal of $N$ in spinorial indices. It defines a positive definite, Hermitan  inner product for spinor fields in $N$, in terms of which the dual of $\lambda^A$ (denoted by $\lambda^{\dagger A}$) is given by
\begin{eqnarray}
\lambda ^{\dagger A}=\sqrt{2}\, \tau^{AA'}\lambda_{A'}\,.
\end{eqnarray}
A spinor norm
\begin{eqnarray}
\label{norm}
\varphi=\lambda_A\lambda ^{\dagger A}
\end{eqnarray}
may then be defined. 
Let $D_{AB}$ be the the spin connection of $(N,h_{ab})$ \cite{sen_quantum_1982} and consider a spinor field subject to the Dirac equation 
\begin{eqnarray}
\label{sw}
{D}_{AC}\lambda ^C=0\,.
\end{eqnarray}
As (\ref{sw}) is an elliptic system, a spinor field $\lambda^A$ that satisfies (\ref{sw}) is subject to a certain boundary condition. Near infinity, we require $\lambda^A=\lambda_{\,0}^A+\mathcal{O}(1/r)$ where $\lambda_{\,0}^A$ is a covariantly constant spinor defined with respect to the flat connection of $\eta_{ab}$. When the inner boundary is non-empty and consists of  a smooth, marginally trapped spherical surface $S$ (or possibly a finite disjoint union of them), we impose the APS boundary condition \cite{atiyah1975spectral} (see also \cite{herzlich1997penrose}) on $\lambda_A$.  With the two dimensional Dirac operator of $S$ given by   
\begin{eqnarray}\label{2d}
\cancel\partial_{A}{}^C=
    \begin{pmatrix}
        0 & m^a \cancel\partial_a \\
    	-\bar m^a \cancel\partial_a & 0
    \end{pmatrix}\,,
\end{eqnarray}
where $\cancel\partial_a$ is the metric connection of $S$ and $(m^a,\bar m^a)$ are the standard complex null tangents of $S$. $\lambda^A$ is said to satisfy the APS (spectral) boundary condition in that
\begin{eqnarray}\label{aps1}
\lambda_A=\sum_{n=0}^{\infty}\,, a_n\lambda_{nA},\quad
a_n\in \mathbb{C}\,,
\end{eqnarray}
and $\lambda_{nA}$ are eigenspinors given by
\begin{eqnarray}\label{aps2}
\cancel\partial_{A}{}^C\lambda_{nC}=-\mu_n\lambda_{nA}\,, \quad\mu_n>0
\quad\hbox {for}\,\, n=0,1,2\dots
\end{eqnarray}
and $\{\lambda_n^A\}$ constitutes an orthonormal basis in the sense that
\begin{eqnarray}\label{aps3}
\int_S\, \lambda_{mC}\lambda_{n}^{\dag C}=\delta_{mn}\,.
\end{eqnarray}
The APS boundary condition means that $\lambda_A$ is spanned linearly by the eignespinors of $\cancel\partial_{A}{}^C$ with negative eigenvalues. 

Throughout the present work, the notations for two spinors will follow that in \cite{penrose_spinors_1984} unless otherwise stated. Contraction of tensorial and spinorial indices are always defined with respect to $h_{ab}$ and the symplectic form $\epsilon_{AB}$ respectively.

\section{Angular velocity of a  rotating black hole and momentum constraint}\label{sec:ang}

One way to understand the spinor approach to the positive energy theorem is that a spinor together with its dual defined in terms of the  timelike unit normal to an initial data set generate a spin dyad and in turn a Newman-Penrose tetrad to parametrise the Hamiltonian of the gravitational field of the initial data.
With this physical picture in mind, the first obstacle to be overcome in the present context is to define an appropriate shift vector in the Hamiltonian of the gravitational field to describe the rotation of black hole. In the case of the Kerr metric, the angular velocity is defined in terms of the stationary structure of spacetime. In our case, we need to define the angular velocity without any reference to a  Killing vector field.

\subsection{Angular Velocity and momentum constraint}

To see the way ahead, consider the following  vector field
\begin{eqnarray}
    V_{a}=K_{ab}\eta^{b}\,,
\end{eqnarray}
where $\eta^{a}=X^{-1/2}\phi^{a}$.
In terms of $V_a$, the momentum constraint in (\ref{Constraint_Equations_1})  may be rewritten as 
\begin{eqnarray}
\label{continuity0}
\cancel\nabla^a V_a=0\,.
\end{eqnarray}
(\ref{continuity0}) resembles the equation of continuity for a fluid flow on the two plane $\Sigma$ with $V_a$ as the two velocity components of the flow. This prompts us to further calculate $\cancel{\nabla}_{[a}V_{b]}$.
The initial guess is that $\cancel{\nabla}_{[a}V_{b]}$ vanishes and we will have an incompressible fluid flow picture tied to the holomorphic geometry of $\Sigma$. However, this guess turns out not to be quite right but not far off the mark. Instead we have a quasi-conformal Beltrami flow. 

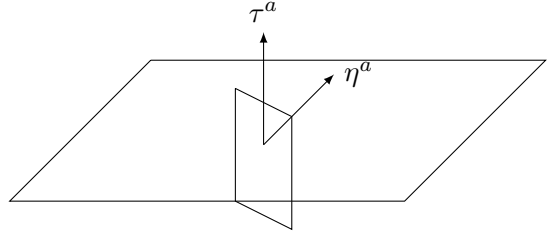
\begin{figure}[h]
    \centering
    \begin{tikzpicture}[scale=0.75]
        \draw (-3,0)--(-5.5,-2.5)--(1.5,-2.5)--(4,0)--(-3,0); 
        \draw (-1.5,-0.5)--(-1.5,-2.5)--(-0.5,-3)--(-0.5,-1)--(-1.5,-0.5); 
        \draw[-latex] (-1,-1.5)--(-1,0.5); 
        \draw (-1,0.5)coordinate (tau)node[above] {$\tau^a$};
        \draw[-latex] (-1,-1.5)--(0.25,-0.25); 
        \draw (0.25,-0.25)coordinate (alpha)node[right] {$\eta^a$};
    \end{tikzpicture}
    \caption{Newman-Penrose tetrad adapted to the geometry of $\Sigma$. }\label{fig2}
\end{figure}

To see this, we will employ the Newman-Penrose formalism (moving frame) approach to compute 
$\cancel{\nabla}_{[a}V_{b]}$. The employment of the NP formalism is convenient but not essential.  A concrete set of NP tetrad adapted to the geometry of $\Sigma$ may be defined as follows (see Fig.~\ref{fig2}):
\begin{eqnarray}\label{eq:NP}
l^a=\frac{1}{\sqrt{2}}(\tau^a+\eta^a),&\ \ \ \ \   n^a=\frac{1}{\sqrt{2}}(\tau^a-\eta^a), \nonumber\\
m^a=\frac{1}{\sqrt{2}}(X^a+iY^a),&\ \ \ \ \ \bar{m}^a=\frac{1}{\sqrt{2}}(X^a-iY^a),
\end{eqnarray}

\noindent where $\tau^a$ is the unit normal vector to the initial slice $S$, and $m^a$, $\bar m^a$ are two complex tangents of $\Sigma$ orthogonal to $\tau^a$ and $\eta^{a}$. We may  easily verify that 
$$l^an_a=1,\ m^a\bar{m}_a=-1,\ l^a m_a=l^a \bar m_a=n^am_a=n^a\bar m_a=0.$$
It is then sufficient to calculate
\begin{eqnarray}\label{eq:cur}
m^a\bar{m}^b \cancel{\nabla}_{[a}V_{b]}&=&m^a\bar{m}^b{P_{a}}^{c}{P_{b}}^{d} D_{[c}(K_{d]e}\eta^{e})\nonumber\\
&=&m^a\bar{m}^b\eta^c(\nabla_a\nabla_b\tau_c-\nabla_b\nabla_a\tau_c)\nonumber\\
&=&(\Psi_2-\bar{\Psi}_2)|_{\Sigma}\,.
\end{eqnarray}
In the first equality of the above expression, the second term on the right hand side vanishes because $\eta^{a}$ is hypersurface orthogonal and the term of the third vanishes owing to $m^{a}\eta_{a}=\bar{m}^a\eta_{a}=0$. $"|_{\Sigma}"$ means retriction to the two-plane $\Sigma$. But in the rest part of this section, we omit this symbol for the sake of simplicity.
By one of the NP structure equation, we have
\begin{eqnarray} \label{NP}
-D\epsilon'-D'\epsilon&=&(\tau-\bar{\tau}')\alpha+(\bar{\tau}-\tau')\beta+(\epsilon+\bar{\epsilon})\epsilon'\nonumber\\
&&+(\epsilon'+\bar{\epsilon'})\epsilon-\tau\tau'+\kappa'\kappa+\Psi_2\,.
\end{eqnarray}
To further express $\Psi_2-\bar{\Psi}_2$ in terms of spin coefficients,  in what follows we shall establish that 
 $\Psi_2=-\kappa\kappa'$. 

To this end, we start with 
\begin{eqnarray}
\epsilon+\bar{\epsilon}&=&l^a\nabla_a(\log X^\frac{1}{2})\nonumber\\
&=& \frac{1}{\sqrt 2}(\tau^a+\eta^a)(\nabla_a \log X^\frac{1}{2})\label{epsilon-0}\,.
\end{eqnarray}
Because $\phi^a$ is a Killing vector, it is obvious that $\eta^a\nabla_a X=0$. Further, as $X$ is intrinsically defined in $N$ and does not dependent on the time evolution of the initial
data set, we necessarily have $\tau^a\nabla_a X=0$. Togther these imply from (\ref{epsilon-0}) that
\begin{eqnarray}\label{eq:epsilon}
\epsilon+\bar{\epsilon}=0\,.
\end{eqnarray} 
Apart from that, we have
\begin{align}
    \bar{\tau}-\tau'&=-l^{a}n^{b}\nabla_{b}\bar{m}_{a}+n^{a}l^{b}\nabla_{a}\bar{m}_{b}\nonumber\\
    &=-2l^{a}n^{b}\nabla_{[b}\bar{m}_{a]}=-2l^{a}n^{b}\partial_{[b}\bar{m}_{a]}=0\,,\label{eq:tau}\\
	\epsilon+\epsilon'&=\frac{1}{2}n^{b}l^{a}\nabla_{a}l_{b}-\frac{1}{2}n^{b}n^{a}\nabla_{a}l_{b}-\frac{1}{2}\bar{m}^{b}l^{a}\nabla_{a}m_{b}+\frac{1}{2}\bar{m}^{b}n^{a}\nabla_{a}m_{b}\nonumber\\
	&=X^{-1/2}n_{b}l^{a}\nabla_{a}\phi^{b}-X^{-1/2}\bar{m}_{b}m^{a}\nabla_{a}\phi^{b}=0\,.\label{epsilongamma}
\end{align}
From the vanishing results of (\ref{eq:epsilon}) and (\ref{epsilongamma}), we have 
\begin{eqnarray}\label{eq:gammaminusepsilon}
	D\epsilon'-D'\epsilon=D\epsilon-D'\epsilon=\sqrt{2}\eta^{a}\nabla_{a}\epsilon=0\,.
\end{eqnarray}
From (\ref{eq:tau}), (\ref{epsilongamma}) and  (\ref{eq:gammaminusepsilon}), (\ref{NP}) may be written as
\begin{eqnarray}
\Psi_2-\bar{\Psi}_{2}=-\kappa\kappa'+\bar{\kappa}\bar{\kappa}'\,.
\end{eqnarray}
 (\ref{eq:cur}) then takes the form
\begin{eqnarray}
m^a\bar{m}^b\cancel\nabla_{[a}V_{b]}&=(\Psi_2-\bar{\Psi}_2)\nonumber\\
&=(-\kappa\kappa'+\bar{\kappa}\bar{\kappa}')\,.
\end{eqnarray}
From the definitions of spin coefficients, $\kappa\kappa'$ is further given by
\begin{eqnarray}
\kappa\kappa'=&&m^a\tau^b\nabla_b\tau_a\cdot \bar{m}^a\tau^b\nabla_b\tau_a-4m^a V_a\cdot \bar{m}^a V_a\nonumber\\
&&+m^a\nabla_a\log X^\frac{1}{2}\cdot \bar{m}^a\nabla_a\log X^\frac{1}{2}-2m^a\tau^b\nabla_b\tau_a\cdot \bar{m}^a v_a\nonumber\\
&&+2\bar{m}^a\tau^b\nabla_b\tau_a\cdot m^a V_a-m^a\tau^b\nabla_b\tau_a\cdot \bar{m}^a\nabla_a\log X^\frac{1}{2}\nonumber\\
&&-\bar{m}^a\tau^b\nabla_b\tau_a\cdot m^a\nabla_a\log X^\frac{1}{2}-2m^a V_a\cdot \bar{m}^a\nabla_a\log X^\frac{1}{2}\nonumber\\
&&+2\bar{m}^a V_a\cdot m^a\nabla_a\log X^\frac{1}{2}\,.
\end{eqnarray}
Therefore we have
\begin{eqnarray}
\kappa\kappa'-\bar{\kappa}\bar{\kappa}'=4m^a\cancel\nabla_a\log(N X^{-\frac{1}{2}})\cdot \bar{m}^a V_a-4\bar{m}^a\cancel\nabla_a\log(N X^{-\frac{1}{2}})\cdot m^a V_a\,,
\end{eqnarray}
where $N$ is the lapse function of the initial data set  which is supposed to be freely specified, and given by
\begin{eqnarray}
\tau^b\nabla_b\tau_a=\nabla_a\log N\,.
\end{eqnarray}
As $m^{[a}\bar{m}^{b]}$ is the basis tensor of the antisymmetric tensor space of surface $\Sigma$, we  conclude that
\begin{eqnarray}
\cancel\nabla_{[a}V_{b]}=V_a\cancel\nabla_b\log (N X^{-\frac{1}{2}})-V_b\cancel\nabla_a\log (N X^{-\frac{1}{2}})\,.
\end{eqnarray}
Then we have
\begin{eqnarray}\label{curl}
\cancel\nabla_{[a}\tilde{V}_{b]}=0\,,\quad {\rm on}\ \Sigma
\end{eqnarray}
where
\begin{eqnarray}\label{vtilde}
\tilde{ V}_{b}=\frac{1}{2}N X^{-\frac{1}{2}}V_b\,.
\end{eqnarray}
Given (\ref{curl}), by  the Poincare Lemma,  there exists a scalar function $\Omega$ such that
\begin{eqnarray}\label{potentialomega}
V_a=\frac{1}{2}N^{-1}X^{\frac{1}{2}}\partial_a \Omega\,.
\end{eqnarray}
$\Omega$ is determined up to a constant by (\ref{potentialomega}). Given the asymptotic falloff of $K_{ab}\sim O(1/r^2)$ together with  (\ref{potentialomega}),  
\begin{eqnarray}\Omega\sim \mathcal{O}(1/r^3)\,,\label{asymptotic}\end{eqnarray}
as $r\rightarrow \infty$
and we may fix this constant to be zero. Further, from the definition of $\Omega$ in terms of $K_{ab}$ and lapse dependent, it is defined on an initial data set and dependent on the 3+1 decomposition of spacetime, unlike the case of the Kerr metric. 
	\begin{definition}
	Let $(N, h_{ab},K_{ab})$ be a quasi-stationary initial data set as defined in the preceding section. Define the angular velocity of this initial data set 
 as the potential $\Omega$  as presented in (\ref{potentialomega}).
\end{definition}

As an example of the above definition, the angular velocity of a Kerr black hole  defined by  (\ref{potentialomega}) is given in the Boyer-Linquist coordinates as 
\begin{eqnarray}
	\Omega=\frac{2mar}{(r^2+a^2)^2-(r^2-2mr+a^2)a^2\sin^2\theta}\sim\mathcal{O}({1/r^3})\,,\,\hbox{when}\,\, r\rightarrow\infty\,.
	\end{eqnarray}
\vskip 10pt
\subsection{Compressible fluid flow and momentum constraint}

With the  angular velocity $\Omega$ defined in the preceding subsection, a  fluid dynamic picture for the 
momentum constraint emerges naturally. 

To see this, in terms of the flat connection of $\Sigma$ and $v_a=XV_a$, (\ref{continuity0})  and (\ref{curl}) may be expressed respectively as  
\begin{eqnarray}\label{continuity}
    \partial^{a}v_{a}=0\,,
\end{eqnarray}
and 
\begin{eqnarray}
     \partial_{[a} N X^{-\frac{3}{2} }v_{b]}=0\,.\label{continuity2}
\end{eqnarray}
(\ref{continuity}) resembles  the equation of continuity for a compressible fluid flow in the two plane $\Sigma$ whose velocity field given by $v_a$. 


The close analogy with a two-dimensional steady compressible fluid flow also suggests to us to define a stream function $\Psi$ in terms of $\Omega$ as     
\begin{eqnarray}\label{rela-two-poten}
\partial_{a}\Omega=w\,\epsilon_{ab}\partial^{c}\Psi\,,
\end{eqnarray}
where $w=\frac{1}{2}NX^{-1}$, $\epsilon_{ab}$ is the two-dimensional Levi-Civita symbol. 
Written in terms of the Euclidean coordinates $(x,y)$ pertained to the flat connection of $\Sigma$, (\ref{rela-two-poten}) takes the form 
\begin{eqnarray}
    	w\,\Omega_{x}=\Psi_{y}\,,\ \ \ w\,\Omega_{y}=-\Psi_{x}\,,\nonumber
\end{eqnarray}
which is the familiar quasi-conformal map. 
 $\Omega$ and $\Psi$ satisfy 
\begin{eqnarray}
    (w\,\Omega_{x})_{x}+(w\,\Omega_{y})_{y}=0\,,\label{Omega-eq}\\
     (w\,\Psi_{x})_{x}+(w\,\Psi_{y})_{y}=0\,.\nonumber
	\end{eqnarray}
As in a two-dimensional fluid flow, $\slashed\nabla_a\Psi$ is tangent to the flow line which are level sets of of $\Omega$ while $\slashed\nabla_a\Omega$
is orthogonal to it. 
 
\begin{figure}[htbp]
	\centering	\includegraphics[width=1.0\textwidth]{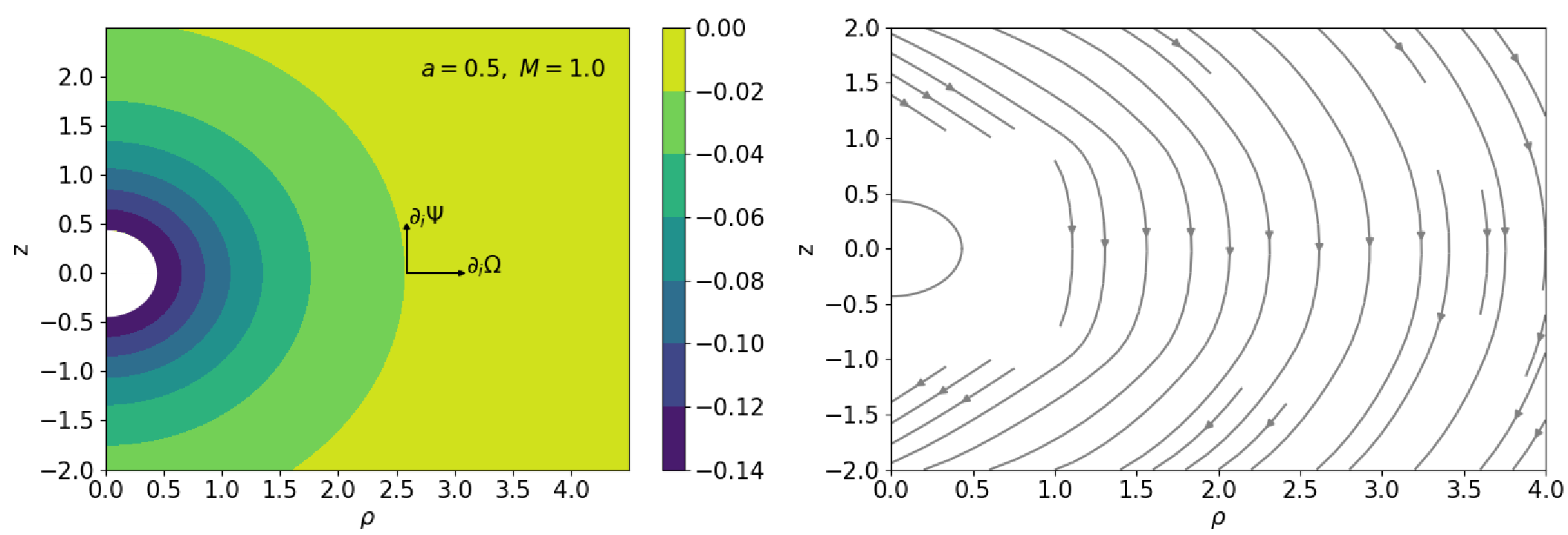}
	\caption{The left panel is the level sets of $\Omega$ for the Kerr metric at $a=0.9$. The right panel is the streamline flow of $\nabla_a \Psi$ at $a=0.9$.}\label{fig:fluid}
\end{figure}

In the case of the Kerr metric, the explicit form of $\Psi$ was first  written down in \cite{dain2006variational} without being aware of the fluid dynamic correspondence. It is stated as 
 \begin{eqnarray}
    	\Psi=2ma(\cos^3\theta-2\cos\theta)-\frac{2ma^3\cos\theta\sin^4\theta}{r^2+a^2\cos^2\theta}\,.     \nonumber
\end{eqnarray}
 See Fig.~\ref{fig:fluid} for an illustration of the fluid flow picture for the case of the Kerr metric near the horizon. 

\subsection{Level sets of $\Omega $ and shearfree structure of the horizon}

\begin{figure}
    \centering
    \begin{tikzpicture}[scale=0.5]
        \draw (0,2.4) ellipse (3.6 and 4.8);
        \draw (0,-2.4) arc (-90:90:2.4 and 4.8);
        \draw (1.7,5)coordinate (gamma)node[below left] {$\gamma$};
        \draw[-latex] (2.4,2.4) -- (3.4,2.4);
        \draw (3.4,2.4)coordinate (eta)node[above] {$\eta^a$};
        \draw[-latex] (2.4,2.4) -- (2.4,3.6);
        \draw (2.4,3.6)coordinate (s)node[above right] {$s^a$};
        \draw[-latex] (2.4,2.4) -- (3.2,1.6);
        \draw (3.2,1.6)coordinate (nu)node[below] {$\nu^a$};
    \end{tikzpicture}
    \caption{Axisymmetric trapped surface.}\label{Axisymmetric structure of a trapped surface.}
\end{figure}

The next question to be addressed is whether $\Omega$ is constant at the horizon, as in the case of the Kerr metric. Let $\gamma$ be the generating curve of the axisymmetric horizon, $s^a$ and $\nu^a$ are respectively the tangent and normal to $\gamma$ (see Fig.~\ref{Axisymmetric structure of a trapped surface.}).  
We need to prove that $s^a \nabla_a \Omega = 0$. On account of (\ref{potentialomega}), this will follow provided $s^a \eta^b K_{ab} = 0$ on $S$. 

By Lie dragging of $s^a$ and $\nu_a$ along the circular integral curves of $\phi^a$ by the normalised vector field $\eta^a$,  the vector fields are extended to the entire trapped surface. Define a NP tetrad adapted naturally to the geometry of the horizon as follows. 
\begin{eqnarray}\label{Newmann_Penrose_2}
L^a = \frac{1}{\sqrt{2}} (\tau^a + \nu^a)\,,& N^a = \frac{1}{\sqrt{2}} (\tau^a - \nu^a)\,, \\
M^a = \frac{1}{\sqrt{2}} (s^a + i \eta^a)\,,& \bar{M}^a = \frac{1}{\sqrt{2}} (s^a - i \eta^a)\,.
\end{eqnarray}
Then $s^a \eta^b K_{ab} = -i (K_{ab} M^a M^b - K_{ab} \bar{M}^a \bar{M}^b) = -i (\sigma - \bar{\sigma} + \bar{\sigma}' - \sigma')$. It is sufficient
to prove that $\sigma=\sigma'=0$ at the horizon and the proof comes down to the computation of certain spin coefficients. 

By construction, the two null vectors $L^{a}$ and $N^{a}$ are orthogonal to the 2-surface $S$ and therefore  $\rho$ and $\rho'$ are both real at surface $S$ \cite{penrose_spinors_1984}. Together with the null character of  $L^{a}$ and $N^{a}$, we also have \cite{penrose1984spinors}
\begin{eqnarray}
    \kappa=\kappa'=0\,.
    \label{kappa}
\end{eqnarray}

\noindent Consider the future trapped case in which $\rho=0$. The reality of $\rho$ and $\rho'$ enables us to write  
\begin{eqnarray}
    \rho=-\gamma^{ab}K_{ab}- p\,,\\
    \rho'=-\gamma^{ab}K_{ab}+ p\,,
\end{eqnarray}
where $\gamma_{ab}$ and $p$ are respectively the two metric of $S$ and $p$  the mean curvature of $S$. The future marginally trapped condition together with quasi-stationary condition then imply that 
\begin{eqnarray}
\label{rho}
    \rho=\rho'=p=0\,.
\end{eqnarray}
Next, due to the vanishing of the mean curvature $p$ given in (\ref{rho}), we shall show that 
\begin{eqnarray}
\label{shear}
    \sigma=\bar\sigma'\,.
\end{eqnarray}

To see this,
\begin{eqnarray}
        p&=&0\nonumber\\
    &=&\gamma^{ab}\nabla_a\nu_b\nonumber\\
    &=&(M^a\bar M^b+\bar M^aM^b)\nabla_a\nu_b\nonumber\\
    &=&\frac{1}{2}(s^as^b\nabla_a\nu_b+\eta^a\eta^b\nabla\nu_b)\,.\label{mean}
\end{eqnarray}
where the last inequality follows from the definition of the NP tetrad given in (\ref{Newmann_Penrose_2}). By construction, the Lie dragging of $\nu^a$
along the circular integral curves of $\phi^a$ by $\eta^a$ enables us to infer
\begin{eqnarray}
\eta^aD_a\nu^b=\nu^aD_a\eta^b\,.\label{lie2}
\end{eqnarray}
Together with $\eta^a\eta_a=1$, we may deduce from (\ref{lie2}) that  
\begin{eqnarray}
    \eta^a\eta^b\nabla_b\nu_a=\eta^b\nu^a\nabla_a\eta_b=0\,. \label{etaa}
\end{eqnarray}
It then follows from  (\ref{mean}) and (\ref{etaa})  that 
\begin{eqnarray}
    s^as^b\nabla_a\nu_b=0\,.\label{geodetic}
\end{eqnarray}
From (\ref{Newmann_Penrose_2}), 
\begin{eqnarray}
    \sigma&=&M^aM^b\nabla_b(\tau_a+\nu_a)\nonumber\\
    &=&\frac{1}{2\sqrt 2}(K_{ab}M^aM^b+M^aM^b\nabla_b\nu_a)\nonumber\\
    &=&\frac{1}{2\sqrt 2}(K_{ab}M^aM^b+(s^a+i\eta^a)(s^b+i\eta^b)\nabla_b\nu_a)\,.\label{shear-1}
\end{eqnarray}
Further,
\begin{eqnarray}
    &&(s^a+i\eta^a)(s^b+i\eta^b)\nabla_b\nu_a\nonumber\\
    &=&(s^as^b\nabla_b\nu_a-\eta^a\eta^b\nabla_b\nu_a)+i(s^a\eta^b\nabla_b\nu_a+\eta^as^b\nabla_b\nu_a)\nonumber\\
    &=&i(s^a\eta^b\nabla_b\nu_a+\eta^as^b\nabla_b\nu_a)\,,\label{shear-11}
\end{eqnarray}
where the last equality follows from (\ref{etaa}) and (\ref{geodetic}). Inserting (\ref{shear-11}) back into (\ref{shear-1}), we then find 
\begin{eqnarray}
        \sigma=\frac{1}{2\sqrt 2}\big[K_{ab} M^a M^b+i(s^a\eta^b\nabla_b\nu_a+\eta^as^b\nabla_b\nu_a) \big]\,.\label{shear-12}
\end{eqnarray}
By similiar calculations, we also have 
\begin{eqnarray}
        \sigma'=\frac{1}{2\sqrt 2}\big[K_{ab}\bar M^a\bar M^b-i(s^a\eta^b\nabla_b\nu_a+\eta^as^b\nabla_b\nu_a)\big]\,. \label{shear-2}
\end{eqnarray}
It may be checked from (\ref{shear-12}) and (\ref{shear-2}) that $\sigma=\sigma'$ in (\ref{shear}) is valid. 

Now the vacuum NP structure equations may be given as 
\begin{align}\label{Structure_Equations_0}
D \rho-\delta^{\prime} \kappa&=\rho^{2}+\sigma \bar{\sigma}+\rho(\varepsilon+\bar{\varepsilon})-\bar{\kappa} \tau-\kappa\left(3 \alpha+\bar{\beta}+\tau^{\prime}\right)\,,\\
D'\rho'-\delta \kappa'&=(\rho')^2+\sigma'\bar{\sigma}'-\rho'(\gamma+\bar{\gamma})-\bar{\kappa}'\tau'-\kappa'(\tau-3\beta-\bar{\alpha})\,.\label{eq:NP2}
\end{align}
In view of (\ref{kappa}) and (\ref{rho}),  
 (\ref{Structure_Equations_0}) becomes
\begin{eqnarray}
    D\rho = \sigma\bar{\sigma}\,.\label{stable}
\end{eqnarray}
By one of the hypotheses,  $S$ is  stable embedded in Riemannian manifold $\Sigma$, which gives $D\rho\leq 0$. (We can let $\tau^{a}D_{a}\rho=0$ since all of our works are restricted to the initial slice). Putting this back into (\ref{stable}) enables us to conclude that $\sigma=0$. By (\ref{shear}), we also have $\sigma'=0$. This establishes that $\Omega$ is constant in $S$, as in the case of the Kerr metric. The past trapped case is similiar. 

Since $\Omega$ is governed by the elliptic equation displayed in   (\ref{Omega-eq}), 
subject to the Dirichlet boundary condition that $\Omega$ is a non-zero constant at the horizon together with the asymptotic behaviour prescribed by (\ref{asymptotic}), the maximum principle then implies that $\Omega$ is non-zero on the  $\Sigma$ plane. This also means that $\Omega$ is either strictly positive or negative in $\Sigma$. 
A positive (negative) $\Omega$ may be respectively regarded as a black hole rotating (counter rotating) along the direction of the integral curves of $\phi^a$. 

\section{Geometric inequality as generalised first law of black hole mechanics}\label{sec:theorem}

With a geometric definition of angular velocity of a rotating black hole in place, we are in a position to prove the following. 

\begin{theorem}\label{theorem}
	Let $(N,h_{ab},K_{ab})$ be asymptotically Euclidean quasi-stationary initial data with a multi-connected, axisymmetric inner boundary $S_{1}$$\cdots$$S_{i}$, each connected component is a  stable minimal surface.  Then the ADM mass $M$ satisfies the inequality
	\begin{eqnarray}
		M\geq c\sum_{i}r_{i}+\sum_{i}\Omega_i\,J_i\,,
	\end{eqnarray}
	where $c=\inf_{\cup S_i}\varphi$ and $0<c<1$ (see \ref{eq:infimum_phi}), $r_i$, $\Omega_i$, $J_i$ are respectively the  areal radius, angular velocity and angular momentum of a component of $S_{i}$.
\end{theorem}

\noindent{\bf Remark I}: Though asymptotically $K_{ab}=\mathcal{O}(1/r^2)$, the ADM linear momentum vanishes given that the initial data set is quasi-stationary. 

\noindent{\bf Remark II}: With the assumption that initial slice is quasi-stationary, $S$ is a marginally future (past) trapped surface, then $S$ may be proven to be a minimal surface.

\vskip 10pt
\noindent{\bf Proof}

Let $\lambda_A$ be the spinor field in $N$ that satisfies the Dirac equation (\ref{sw}) and subject to the prescribed boundary conditions at the horizon and near spatial infinity given in Section 2. Consider first the case when the horizon consists of only one single connected component $S$ and denote by $S_\infty$ be the limiting coordinate sphere at infinity. 

We begin with the identity
\begin{eqnarray}\label{norm0}
&\frac{1}{8\pi}\int_{S_\infty}\,(D_a\varphi-K_{ab}N^{b})\nonumber\\
&=\frac{1}{8\pi}\int_{N}(\triangle\varphi-K_{ab}D_a N^{b})+\frac{1}{8\pi}\int_{S}(D_a\varphi -K_{ab}N^{b})d S^a\,,
\end{eqnarray}
which relates  the total Hamiltonian of the gravitational field to the vacuum constraints and the Hamiltonian boundary terms. We choose the lapse function $\varphi$
to be the spinor norm given by 
\begin{eqnarray}
    \varphi=\lambda_A\lambda^{\dagger A}\,,
\end{eqnarray}
and the shift vector to be $N^a=\Omega\phi^a$ which describes the rotation of a black hole, as suggested by the Kerr metric. 

Given the asymptotic behaviour of  $\Omega$ (see (\ref{asymptotic})), the angular part of the Hamiltonian boundary term near infinity vanishes and we are left with the ADM mass $M$ at infinity. (\ref{norm0}) then becomes 
\begin{eqnarray}
   M =\frac{1}{8\pi}\int_{N}(\triangle\varphi-K_{ab}D_aN^{b})+\frac{1}{8\pi}\int_{S}(D_a\varphi -K_{ab}N^{b})d S^a\,.
\label{ham1}
\end{eqnarray}

The proof of the theorem will be divided into two parts, concerning respectively with the ADM mass and angular momentum.

\subsection{ADM mass}\label{sec:ADM}

Suppose the inner boundary of $N$ consists
of a smooth spherical surface $S$ which is marginally trapped.   Let
\begin{eqnarray}
\label{potential} \varphi=\lambda_A\lambda^{\dagger A}\,.
\end{eqnarray}
To evaluate the inner boundary term involving the normal derivative of $\varphi$, sufficiently close to
$S$, $\nu^a$ defines a smooth radial geodesic flow and 
consider
 the moving frame $(m^a, \bar m^a)$ in $S$ with  $m^a,
\bar m^a$  the complex null tangents to $S$ which satisfy $m^a\bar
m_a=1$. A moving frame field in the vicinity of $S$, still denoted
by $(\nu^a, m^a, \bar m^a)$, may then be constructed by the parallel
transport of $(\nu^a, m^a, \bar m^a)$ at $S$ along the radial
geodesics. In terms of this moving frame, (\ref{sw}) may be expressed as
\begin{eqnarray}
\frac{\,1}{\sqrt 2}\left(\partial_\nu+\frac{\,p}{\,2}\right)\lambda_A
=\,-\cancel\partial_{A}{}^C\lambda_C\,, \label{diraceq}
\end{eqnarray}
where $\partial_\nu=\nu^a\partial_a$ and $\cancel\partial_{A}{}^C$ is
as that defined in (\ref{2d}).

Consider the quadratic form
\begin{eqnarray}\label{quad1}
\frac{\,1}{\sqrt 2}\,\,\lambda^{\dag A}
\left(\partial_\nu+\frac{\,p}{\,2}\right)\lambda_A =-\lambda^{\dag
A}\,\cancel\partial_{A}{}^C\lambda_C\,.
\end{eqnarray}
As $\lambda^{\dag}{}_A$ also satisfies (\ref{sw}) and this enables us
to interchange  the roles of $\lambda^{\dag A}$ and $\lambda^{\dag
A}$ in (\ref{quad1}) and we have
\begin{eqnarray}\label{quad2} 
\frac{\,1}{\sqrt
2}\,\,\lambda_{A}\left(\partial_\nu+\frac{\,p}{\,2}\right)\lambda^{\dag
A} =-\lambda_{A}\,\cancel\partial^{AC}\lambda_C^{\dag}\,.
\end{eqnarray}
From the definition of $\varphi$ given in (\ref{potential}), (\ref{quad1}) and (\ref{quad2}) together give
\begin{eqnarray}
\label{dirac2} \frac{\,1}{\sqrt
2}\,\,\frac{\partial\varphi}{\partial\nu}=\,-\,\lambda^{\dag
A}\,\cancel\partial_{A}{}^C\lambda_C \,-\,
\,\lambda_{A}\,\cancel\partial^{A}{}^C\lambda_C^{\dag}\,.
\end{eqnarray}
The mean curvature term drops out in (\ref{dirac2}) as $S$ is a minimal surface.

Given the APS boundary conditions prescribed in (\ref{aps1}) and (\ref{aps2}) and together with (\ref{aps3}), it follows from (\ref{dirac2}) that
\begin{eqnarray}\label{ineq0}
&&\int_{S}\,\,\frac{\partial\varphi}{\partial\nu}\nonumber\\
&=&\sum_{n,m}\, a_n\bar a_m\,\mu_n
\int_{S}\,\,\lambda_{nC}\lambda_{m}^{\dag C} \, +\, a_m\bar
a_n\,\mu_m \int_{S}\,\,\lambda_{mC}\lambda_{n}^{\dag C}\nonumber\\
&\ge& 2\sum_{n}\, |a_n|^2\,\mu_n\,.
\end{eqnarray}
For a smooth spherical two surface, $\mu_n\ge\mu_0\ge\frac{1}{r}$
for all $n\ge 0$ \cite{bar1992lower}, where $r$ is the areal radius defined in terms of the area $A$ of $S$ as $A=4\pi r^2$. We may then further deduce from (\ref{ineq0}) that
\begin{eqnarray}
\,\frac{2}{r}\, \sum_{n}\, |a_n|^2
=\frac{2}{r}\,\int_{S}\, \varphi\,,\label{ineq1}
\end{eqnarray}
where the second equality follows from (\ref{potential}) and
(\ref{aps1}). As a result (\ref{ineq0}) becomes
\begin{eqnarray}
\label{ineqq} \frac{\partial\varphi}{\partial\nu}\ge c\, r\,,
\end{eqnarray}
where
\begin{eqnarray} \label{eq:infimum_phi}
    c=\inf_S\varphi\,.
\end{eqnarray}
Subject to the APS boundary condition, the
zero points of $\lambda^A$ necessarily stay away from the inner
boundary \cite{lau2020spinor} and therefore $c>0$. So (\ref{ineqq}) is a
stronger statement than the positive mass theorem.

When $\partial N$ contains more than one connected components with
$\partial N=\cup_{i=1}^{n} S_i$ for some $n\ge 1$,  inequality (\ref{ineqq}) may be generalised in a straightforward way and becomes
\begin{eqnarray}
\label{ineq}
\int_{S}\,\,\frac{\partial\varphi}{\partial\nu}\ge c\sum_{i=1}^n\,r_i\,,
\end{eqnarray}
where $c=\inf_{\cup S_i}\varphi$ and $r_i$ is the areal radius of the respective $S_i$ for $i=1,..., n$. Subject to the APS boundary condition, $-\cancel{\partial}_A{}^C$
is a positive operator and from which it may further be inferred that $\frac{\partial\varphi}{\partial \nu}>0$
at $S$. The maximum principle together with the asymptotic boundary condition for $\lambda_A$ then imply $0<c\le\varphi<1$ (see \cite{lau2020spinor} for details). In the case of the Schwarzschild metric, $c\rightarrow 0$ when we shrink the radius of the trapped surface and $c$ is not a uniform bound for an arbitrary trapped surface. In the case of a stable minimal surface, 
$c$ is indeed further bounded by certain Sobolev-type constant. This will be presented in another work. In this paper, we will confine our attention to the study of the geometry of rotation of a black hole.

\subsection{Spinor calculations for rotating black holes}

For the inner boundary term $\int_S K_{ab}N^{b}dS^a $, $\Omega$ is constant at $S$ and, given $\phi^a$ is Killing,  $\int_S K_{ab}N^{b}dS^a$ may be identified with the 
Komar integral for angular momentum $J$ and the inner boundary term becomes $\Omega J$. Together with (\ref{ineq}), (\ref{norm0}) may be further written as
\begin{eqnarray}
   M =\frac{1}{8\pi}\int_{N}(\triangle\varphi-K_{ab}D_aN^{b})+cr+\Omega J\,.
\label{ham2}
\end{eqnarray}
To further evaluate (\ref{ham2}), We shall provisionally adopt the assumption that
the spinor norm  is non-zero everywhere in $N$ so that $\varphi>0$. This will be lifted towards the end of the proof. It then makes sense to consider the definition of $K_{ab}$ given by 
\begin{eqnarray}
\label{Kab}
	K_{ab} = -\frac{1}{2 \varphi} \dot{h}_{ab} + \frac{1}{\varphi} D_{(a} N_{b)}\,,
\end{eqnarray}
where $  \dot{h}_{ab} $ is the Lie derivative of the three metric $ {h}_{ab}$ defined with respect to the time evolution vector field of the Hamiltonian. 

It is sufficient to prove the positivity of the volume integral in (\ref{ham1}), in order to establish the inequality stated in Theorem \ref{theorem}. Unlike the standard case of the positive energy proof when the shift vector for translation is constructed in terms of the flagpole of a spinor field, we shall define the shift vector that describes rotation independent of the spinor field and the standard technique in proving the positive energy theorem no longer works. In the first step, through the Gauss-Codazzi equations involving the two plane $\Sigma$ embedded in $N$, what we seek to  do is to relate the angular part in (\ref{norm0}) to the Gaussian curvature of the $\Sigma$ plane and through which contact is made with spinors.

 Combining (\ref{extrinsic}), (\ref{potentialomega}) and (\ref{Kab}), we have  $\dot{h}_{ab}\eta^{a}=0$. We may then infer
\begin{eqnarray}
    K^{ab} D_a N_b =\varphi |K_{ab}|^2\,.
\end{eqnarray}
By the Gauss-Codazzi equation for the embedding of $N$ in spacetime (or the Hamiltonian constraint) together with vacuum and maximal slicing conditions, we have
\begin{eqnarray}
\label{hconstraint}
	R = |K_{ab}|^2\,.
\end{eqnarray}
At the same time, the Gauss-Codazzi equation for the totally geodesic embedding of $\Sigma$ in $N$ gives
\begin{eqnarray}
\label{embed}
	R = 2k\,,
\end{eqnarray}
where $k$ is the Gaussian curvature of the two-plane $\Sigma$.  (\ref{hconstraint}) and  (\ref{embed})  together then imply 
\begin{eqnarray} 
\label{mconstraint}
	 |K_{ab}|^2=2k\,.
\end{eqnarray}
Inserting (\ref{mconstraint}) back into (\ref{ham2}), we then have 
\begin{eqnarray}
   M =\frac{1}{8\pi}\int_{N}(\triangle\varphi-2k\varphi)+cr+\Omega J\,.
\label{ham3}
\end{eqnarray}

Next, by means  of spinor analysis we shall seek  to prove that 
\begin{eqnarray}
\label{lap1}
    \triangle\varphi-2k\varphi\ge 0\,.
\end{eqnarray}
To begin the calculations, we write the three metric as 
\begin{eqnarray}\label{metric}
h_{ab}=\eta_a\eta_b+\gamma_{ab}\,,
\end{eqnarray}
where $\gamma_{ab}$ is the metric on $\Sigma$. By hypothesis, $\phi^a$ is orthogonal to $\Sigma$ and $\phi^a$ is Killing. This implies that $\Sigma$ is totally geodesic in $N$. 
The integrand in  (\ref{lap1}) may be expressed as 
\begin{eqnarray}\label{witten-11}
\triangle\varphi-2k\varphi
=\eta^a\eta^bD_aD_b\,\varphi+\cancel\triangle\varphi-2k\varphi\,,
\end{eqnarray}
where $\cancel\triangle=\gamma^{ab}\cancel\nabla_a\cancel\nabla_a$.
The first term in (\ref{witten-11}) may be further decomposed as 
\begin{align}\label{eq:etaetaDD}
\eta^a\eta^bD_aD_b\,\varphi=2|D_\eta\lambda_A|^2+\lambda^{\dag C}\eta^a\eta^bD_aD_b\lambda_C+\lambda_C\eta^a\eta^bD_aD_b\lambda^{\dag C}\,.
\end{align}
Rewrite (\ref{metric}) as $\eta^a\eta^b=h^{ab}-\gamma^{ab}$, we then have
\begin{align}\label{witten-12}
&\lambda^{\dag C}\eta^a\eta^bD_aD_b\lambda_C+\lambda_C\eta^a\eta^bD_aD_b\lambda^{\dag C}\nonumber\\
&=(\lambda^{\dag C}\triangle \lambda_C\,+\,\lambda_{ C}\triangle \lambda^{\dag C})-\,(\lambda^{\dag C}\cancel\triangle\lambda_{ C}\,+\,\lambda_{C}\cancel\triangle\lambda^{\dag C})
\nonumber\\
&=\frac{R}{2}\varphi\,-\,k\,\varphi\,-
\,2\lambda^{\dagger\,C}{\cancel\nabla}_C{}^N{\cancel\nabla}_N{}^M\lambda_{M}\,-\,
2\lambda_{C}{\cancel\nabla}{}^{\,CN}{\cancel\nabla}_N{}^M\lambda^{\dagger}_{M}\,,
\end{align}
where $R$ and $k$ are respectively the scalar curvature of $N$ and the Gaussian curvature of $\Sigma$. The last equality in (\ref{witten-12}) follows from standard Weitzenbock type identity for the massless Dirac operator $D_A{}^C$ and $\cancel\nabla_A{}^C$. 
Since $\Sigma$ is totally geodesic, the Gauss Codazzi equation of the embedding of $\Sigma$ in $N$ implies that 
\begin{eqnarray}
    \frac{R}{2}-k=0\,,\nonumber
\end{eqnarray}
and (\ref{witten-12}) becomes 
\begin{eqnarray}\label{eq:witten-12-0}
&&\lambda^{\dag C}\eta^a\eta^bD_aD_b\lambda_C+\lambda_C\eta^a\eta^bD_aD_b\lambda^{\dag C}\nonumber\\
&=&(\lambda^{\dag C}\triangle \lambda_C\,+\,\lambda_{ C}\triangle \lambda^{\dag C})-\,(\lambda^{\dag C}\cancel\triangle\lambda_{ C}\,+\,\lambda_{C}\cancel\triangle\lambda^{\dag C})
\nonumber\\
&=&\frac{R}{2}\rvarphi\,-\,k\,\rvarphi\,-
\,2\lambda^{\dagger\,C}{\cancel\nabla}_C{}^N{\cancel\nabla}_N{}^M\lambda_{M}\,-\,
2\lambda_{C}{\cancel\nabla}{}^{\,CN}{\cancel\nabla}_N{}^M\lambda^{\dagger}_{M}\,,
\end{eqnarray}

Further, express the Dirac equation in terms of the moving frame $(\eta^a,m^a,\bar m^a)$ (see (\ref{diraceq}) with $\eta^a$ in place of $\nu^a$) and given that $\Sigma$ is totally geodesic in $N$, we have 
\begin{eqnarray}
\label{equal}
    2|D_\eta\lambda_A|^2=4|\cancel\nabla_{A}{}^C\lambda_C|^2\,.
\end{eqnarray}
Substituting   (\ref{eq:witten-12-0}) and (\ref{equal}) into (\ref{eq:etaetaDD}) and then putting  (\ref{eq:etaetaDD}) back into (\ref{witten-11}), (\ref{witten-11}) takes the form
\begin{eqnarray}
\triangle\rvarphi-2k\rvarphi\nonumber=&\ 4|\cancel\nabla_{A}{}^C\lambda_C|^2\,-\,2\lambda^{\dagger\,C}{\cancel\nabla}_C{}^N{\cancel\nabla}_N{}^M\lambda_{M}\,-\,
2\lambda_{C}{\cancel\nabla}{}^{\,CN}{\cancel\nabla}_N{}^M\lambda^{\dagger}_{M}\nonumber\\
&+\cancel\triangle\rvarphi\,-\,2k\rvarphi\,.\label{witten-12-2}
\end{eqnarray}
At this point, standard spinor identity also gives 
\begin{eqnarray}
\cancel\triangle\varphi=k\varphi+2|\cancel\nabla_{AB}\lambda_C|^2+
2\lambda^{\dagger\,C}{\cancel\nabla}_C{}^N{\cancel\nabla}_N{}^M\lambda_{M}\,+\,
2\lambda_{C}{\cancel\nabla}{}^{\,CN}{\cancel\nabla}_N{}^M\lambda^{\dagger}_{M}\,.\label{identity}
\end{eqnarray}
Putting (\ref{identity})
into  (\ref{witten-12-2}), the second order covariant derivative terms in (\ref{witten-12-2}) and (\ref{identity}) mutually cancel each other and we have 
\begin{eqnarray}
\triangle\varphi-2k\varphi\ge 4|\cancel\nabla_{A}{}^C\lambda_C|^2+2|\cancel\nabla_{AB}\lambda_C|^2-k\varphi\,.
\label{witten-12-1}
\end{eqnarray}
So far we have succeeded in reducing the problem to the two-dimensional spinor geometry in $\Sigma$.  
It is  sufficient to prove that the positive terms  dominate over the Gaussian curvature term in (\ref{witten-12-1}). To this end, we shall further work on (\ref{witten-12-1}) by means of  spinor analysis in $\Sigma$.

Define
\begin{eqnarray}
   \lambda_A=|\lambda|\,o_A\,,
    \label{o}
\end{eqnarray}
where  $|\lambda|=\varphi^{\frac{1}{2}}$. $o_A$ is then a normalisation of $\lambda_A$ with unit spinor norm. The spinor identity in (\ref{identity}) for $o_A$ then gives 
\begin{align}
0&=k+2|\cancel\nabla_{AB}o_C|^2+2o^{\dagger\,C}{\cancel\nabla}_C{}^N{\cancel\nabla}_N{}^M o_{M}\,+\,
2o_{C}{\cancel\nabla}{}^{\,CN}{\cancel\nabla}_N{}^Mo^{\dagger}_{M}\nonumber\\
&=k+2|\cancel\nabla_{AB}o_C|^2-4|\cancel\nabla_{A}{}^{C}o_C|^2\nonumber\\
&\quad + 2{\cancel\nabla}_C{}^N(o^{\dagger\,C}{\cancel\nabla}_N{}^M o_{M})\,+\,
2{\cancel\nabla}{}^{\,CN}(o_{C}{\cancel\nabla}_N{}^Mo^{\dagger}_{M})\,.
\label{identity-1}
\end{align}
For an arbitrary spinor field $\alpha_A$, we have $(\cancel\nabla_{AB}\alpha_C)^\dag=-\cancel\nabla_{AB}\alpha_C^\dag$. With this identity, it may be checked that the divergence terms in (\ref{identity-1}) are purely imaginary and necessarily zero. As a result, (\ref{identity-1}) becomes
\begin{eqnarray}
 4\,|\cancel\nabla_{A}{}^Co_C|^2=k+2\,|\cancel\nabla_{AB}o_C|^2\,.
\label{witten-k}
\end{eqnarray}

On account of the above inequality, it follows from (\ref{witten-k}) that 
\begin{eqnarray}
 k\le 2\,|\cancel\nabla_{AB}o_C|^2\,.
\label{witten-k1}
\end{eqnarray}
Putting (\ref{witten-k1}) back into (\ref{witten-12-2}), we find 
\begin{eqnarray}
\triangle\varphi-2k\varphi\ge 4|\cancel\nabla_{A}{}^C\lambda_C|^2+2|\cancel\nabla_{AB}\lambda_C|^2-2\varphi|\cancel\nabla_{AB}o_C|^2\,.
\label{witten-19}
\end{eqnarray}
In terms of the definition of $o_A$ given in (\ref{o}), we may write 
\begin{eqnarray}
|\cancel\nabla_{AB}\lambda_C|^2=\varphi\,|\cancel\nabla_{AB}o_C|^2 +
|\,\cancel\nabla_{AB}\,|\lambda|\,|^2 \,.
\label{witten-20}
\end{eqnarray}
Substitute (\ref{witten-20}) back into (\ref{witten-19}), we finally have 
\begin{eqnarray}\label{eq:witten-21}
\triangle\varphi-2k\varphi\ge\, 4\,|\cancel\nabla_{A}{}^C\lambda_C|^2+
2\,|\,\cancel\nabla_{AB}\,|\lambda|\,|^2 \,>\,0\,,
\end{eqnarray}
as required. 

\section{Regularisation of  zero points of a spinor field}

So far we have adopted the hypothesis that $\varphi>0$. This means that the spinor field $\lambda_A$ has no zero points in $N$. This requirement may be relaxed by means of regularisation of the zero points in $N$ \cite{lau2020spinor}. Subject to the APS boundary condition at the inner boundary and the asymptotic fall off condition for $\lambda_A$, it may be proved that the
set of zero points lie inside a subset of finite union of compact smooth line segments  $C_k:[0,1] \rightarrow N$ for $k=1,..n$ in $N$. By enclosing the line segments within tubular neighbourhoods $T_k:[0,L_k]\times
D_\epsilon \rightarrow N $  so that $C_k\subset T_k$, $D_\epsilon$
is a geodesic disk of radius $\epsilon$ centered at a point in
$C_k$. inner boundary terms
 $\sum_0^k\int_{\partial T_k}$ apart from the trapped surface $S$. 
By shrinking the radius of the tubes $T_k,k=1,\cdots n $ to a
sufficiently
 small $\epsilon$, it may be checked that  the  positivity argument continues to hold for (\ref{norm0})
  even when   zero points of $\lambda_A$ are taken into consideration. 
  
 Finally we have from (\ref{ham3}) and (\ref{eq:witten-21})  the following inequality
\begin{eqnarray}
M\ge\,c\,r\,+\, \Omega J,\quad\quad 0<c<1.\nonumber
\end{eqnarray}
The generalisation to multi-connected components of a stable marginally trapped surface (future or past) located along the symmetry axis of $\phi^a$ is straightforward. The proof of Theorem \ref{theorem} is then completed.

\section{Concluding Remarks}\label{sec:conclude}
In this work,  a geometric inequality for axisymmetric black hole with angular momentum  is established using spinor approach. This represents a small step forward in our quest to understand the Penrose inequality for rotating black holes. Even for axisymmetric initial data set, many open problems remain before we have a Penrose inequality with angular momentum included. One key question is to have a better estimate of the angular velocity at the horizon in terms of the areal radius. Further relaxation of the maximal slicing condition is also feasible. The present proof relies heavily on the total geodesic structure of the two plane orthogonal to the axisymmetric Killing vector and it is not obvious  how to relax this geometric structure. How far the spinor approach will take us also remains to be understood. 

As far as black hole initial data set is concerned, we have also established the equivalence of the momentum constraint in the axisymmetric initial data set considered here with a conformally invariant  equation for two-dimensional compressible fluid flow. This also suggests a new way to construct new initial data set for binary rotating black holes. This will be further taken up in our future work. 

\section*{Acknowledgments}
YKL is supported by the National Key Research and Development Program of China
 (Grant No. 2021YFC2202501).  Sijie Gao is supported by the NSFC Grants No. 11775022 and 11873044. Discussions and inspirations from the late Sergio Dain is gratefully acknowledged. The long term support to the GR research in the Morningside Center of Mathematics is conducive to the completion of the present work. 

\section*{References}


\bibliographystyle{unsrt}

\bibliography{Geo-Inequa.bib}
\end{document}